\documentclass[%
 aip,
 jcp,%
 amsmath,amssymb,
 reprint,%
]{revtex4-1}

\usepackage{graphicx}
\usepackage{dcolumn}
\usepackage{bm}

\begin{document}

  \title {Mixtures of functionalized colloids on substrates}

  \author{C. S. Dias}
   \email{csdias@cii.fc.ul.pt}
    \affiliation{Centro de F\'isica Te\'orica e Computacional, Universidade de Lisboa, Avenida Professor Gama Pinto 2, P-1649-003 Lisboa, Portugal}

  \author{N. A. M. Ara\'ujo}
   \email{nuno@ethz.ch}
   \affiliation{Computational Physics for Engineering Materials, IfB, ETH Zurich, Wolfgang-Pauli-Strasse 27, CH-8093 Zurich, Switzerland}

  \author{M. M. Telo da Gama}
   \email{margarid@cii.fc.ul.pt}
    \affiliation{Centro de F\'isica Te\'orica e Computacional, Universidade de Lisboa, Avenida Professor Gama Pinto 2, P-1649-003 Lisboa, Portugal}
    \affiliation{Departamento de F\'{\i}sica, Faculdade de Ci\^{e}ncias da Universidade de Lisboa, P-1749-016 Lisboa, Portugal}

  \begin{abstract}
Patchy particles are a class of colloids with functionalized surfaces. Through surface functionalization, the strength and directionality of the colloidal interactions 
are tunable allowing control over coordination of the particle. Exquisite equilibrium phase diagrams of mixtures of coordination two and three have been reported. 
However, the kinetics of self-organization and the feasibility of the predicted structures are still largely unexplored. 
Here, we study the irreversible aggregation of these mixtures on a substrate, for different fractions of two-patch particles. Two mechanisms of mass transport 
are compared: diffusion and advection. In the diffusive case, an optimal fraction is found that maximizes the density of the aggregate. By contrast, for 
advective transport, the density decreases monotonically with the fraction of two-patch colloids, in line with the behavior of the liquid density on the spinodal 
of the equilibrium phase diagram.
  \end{abstract}
  \maketitle

\section{Introduction}

The functionalization of colloids allows fine tuning the directionality of the interactions, an important tool for material 
design.\cite{Ruzicka2011,Glotzer2010,Chen2011,Romano2011a,Kraft2012,Blaaderen2006,Bianchi2011,Pawar2010,Kretzschmar2011,Grzelczak2010}
Recently, there has been a significant advance on both the production of these particles\cite{Sacanna2011,Wilner2012,
Hu2012,Duguet2011,Pawar2008,Shum2010,Wang2012,He2012} and the theoretical understanding of their equilibrium phase diagrams.
\cite{Glotzer2004, Doye2007, Sciortino2010, Sciortino2011, Rosenthal2011, Ruzicka2011,Romano2012, Zaccarelli2005,Zaccarelli2006,Bianchi2006,Russo2010,DelasHeras2012,Varrato2012} 
However, the kinetics of aggregation and self-organization are still largely unexplored.

Pioneer studies of aggregation of patchy colloids in the limit of irreversible binding have revealed non-equilibrium structures which significantly differ from 
the thermodynamic ones.\cite{Dias2013,Dias2013a,Vasilyev2013} However, while in equilibrium the obtained structures do not depend on the history of the assembly process, 
under non-equilibrium conditions this is not the case. The formation of strong bonds between particles, characterized by very long relaxation times,
yields huge energy barriers for the relaxation towards equilibrium. Consequently, kinetically trapped structures are obtained 
which imprint the kinetic pathways of aggregation. Here, we compare two limiting cases of mass transport in solution: advection and diffusion.

The number of bonds that a patchy colloid can form controls the phase diagram of the assembly.
\cite{Zaccarelli2005,Bianchi2006,Bianchi2008,Russo2009,Ruzicka2011,Lu2008} Mixtures of patchy colloids
add novel features to the behavior of these complex fluids, e.g., the formation of interpenetrating gels or bigels.\cite{DelasHeras2012,Varrato2012} 
In particular, mixtures of two- and three-patch
colloids have been considered as they allow control of the ratio between chains and branches, through the fraction of two- and three-patch particles, respectively.
Theoretical equilibrium phase diagrams of these mixtures reveal a drastic reduction of the phase separation region and the possibility of equilibrium low density 
liquids.\cite{Bianchi2006,DelasHeras2011,DelasHeras2011a,Rovigatti2013}

The interest in the aggregation on substrates is twofold. First, from the practical point of view, a substrate works as a nucleation center for growth 
improving the controllability over assembly.\cite{Dias2013,Dias2013a,Gnan2012,Bernardino2012} Second, with substrates it is possible to define a growth direction 
(away from the substrate) and characterize the time evolution of the structure. This possibility constitutes a powerful tool for a systematic theoretical study of 
non-equilibrium growth \cite{Einstein2010}.

We performed extensive kinetic Monte Carlo simulations of the irreversible aggregation on a substrate in the limit of two mechanisms of mass transport: 
diffusion and advection. We show that in the limit of diffusive transport, an optimal ratio of two- and three-patch colloids is found that maximizes 
the density of the structure. By contrast, under advection, the density decreases monotonically with the fraction of two-patch colloids.

This article is organized as follows: In Sec.~\ref{sec.model} we introduce the model and the underlying physical considerations, 
as well as the simulation details; In Sec.~\ref{sec.results} the results for solutions of three-patch colloids (\ref{sec.three})
and mixtures of two- and three-patch colloids (\ref{sec.two-tree}) are presented. Finally, in Sec.~\ref{sec.conclusions} we draw some
conclusions.

\section{Model and Methods}\label{sec.model}

\begin{figure}
  \begin{center}
  \includegraphics[width=\columnwidth]{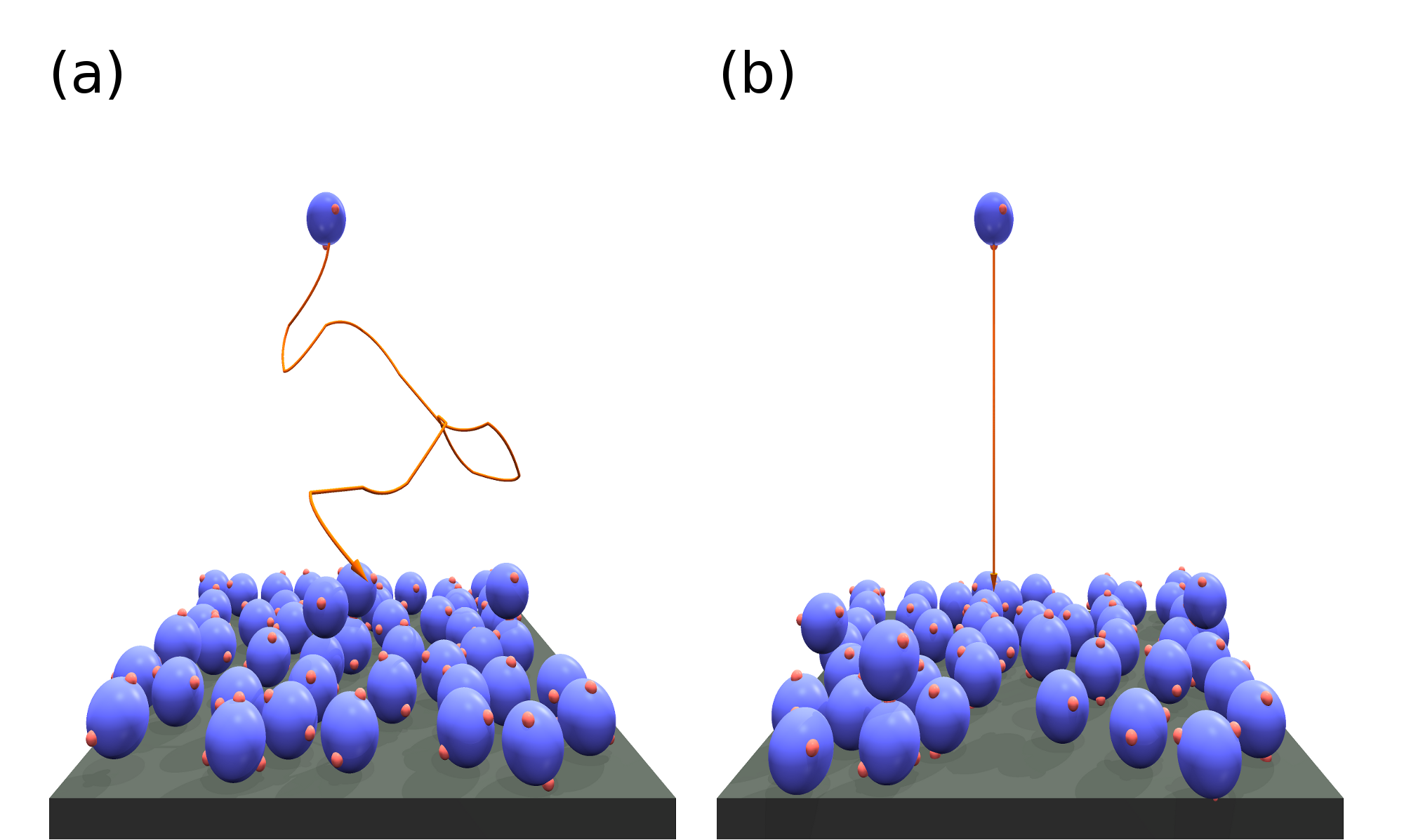} \\
  \end{center}
  \caption{(color online) Schematic representation of the two mechanisms of mass transport towards the substrate: (a) diffusion and (b) advection. 
(a) Brownian motion of a colloid in solution where the particle moves under successive collisions with the solvent and. At each collision, a
 new velocity is generated from the Maxwell-Boltzmann distribution.
(b) Advection is represented by a ballistic motion towards the substrate, following a straight vertical trajectory.
   \label{fig::brownVsball}
   }
\end{figure}

We use the model proposed in Ref.~\cite{Dias2013}. Colloids are considered spherical particles with $n$ patches equally spaced on their surface. 
Interactions are assumed pairwise, repulsive between the spherical cores and attractive between patches. 
In both cases the interaction is short ranged, such 
that the colloid-colloid interaction can be described as excluded volume interaction and only two, almost overlapping, patches can bind. 
Below, we describe in detail the patch-patch interaction as well as the two limits of mass transport in solution: diffusion and advection (see Fig.~\ref{fig::brownVsball}).

\subsection{Patch-patch Interaction}\label{sec.interaction}

\begin{figure}
  \begin{center}
  \includegraphics[width=0.7\columnwidth]{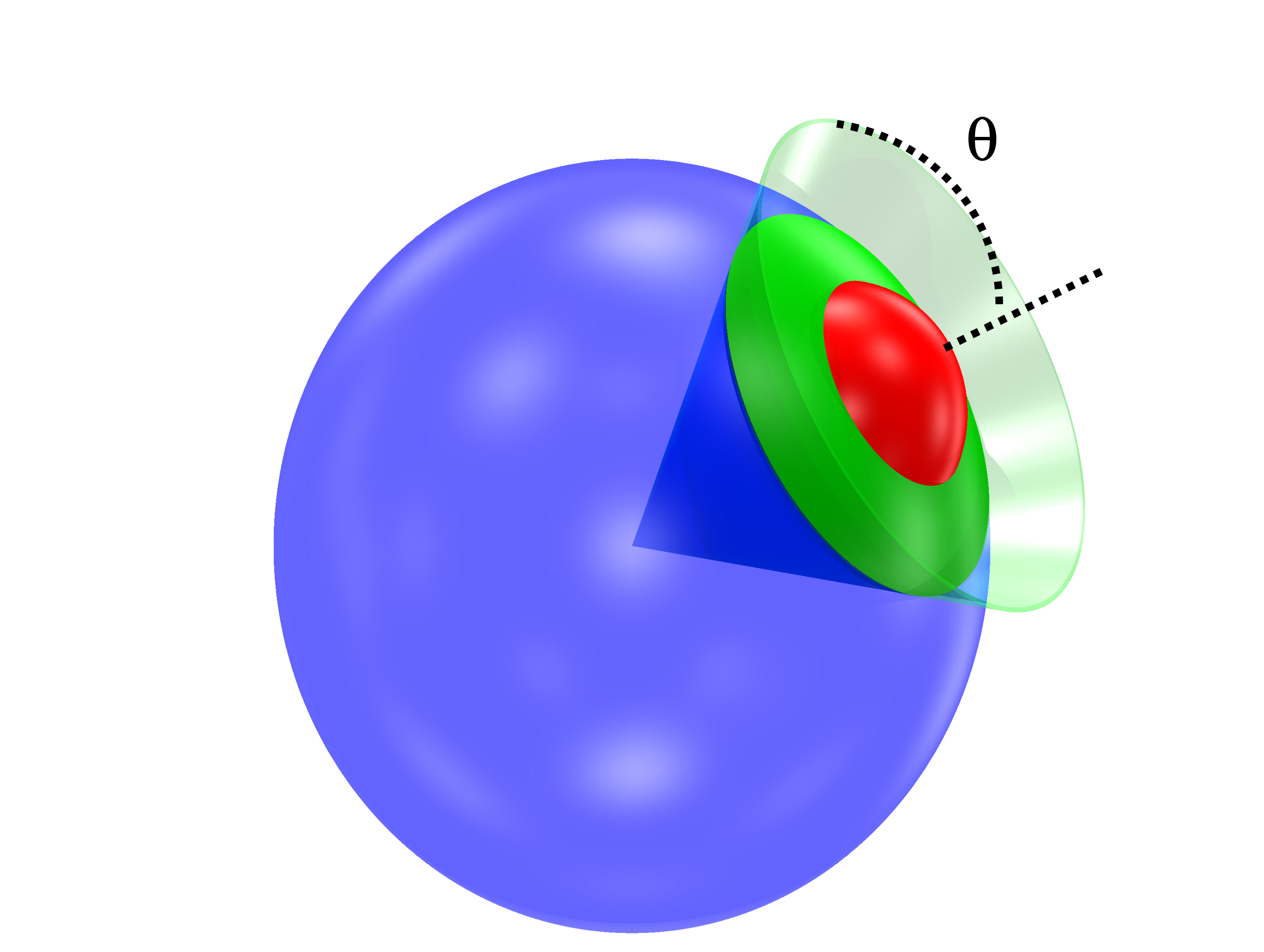} \\
  \end{center}
  \caption{(color online) Schematic representation of a patch (red) on the surface of a
colloid (blue) and its interaction range (green). The limits of the
interaction range are defined by the angle $\theta$ with the center of
the patch.
   \label{fig::interaction.range}
   }
\end{figure}

The patch-patch interaction is considered short range. Thus, we assume that patches of two different colloids can only bind upon collision between colloids. 
For each patch, we define the interaction range on the surface of the colloid, around the patch, describing the area within which the patch-patch interaction is effective. 
As shown in Fig.~\ref{fig::interaction.range}, we define the interaction range (green) of the patch (red) as a circular region on the 
surface of the colloid, parametrized by the angle $\theta$. If the contact point upon collision is within the interaction ranges of both patches, binding 
is successful, otherwise an elastic collision occurs. In the simulation, as only one colloid moves at a time, collision occurs between the moving colloid 
and one colloid bound to the growing structure. In this case, binding occurs with probability $p$ when 
the new arriving colloid collides within one of the interaction ranges of the other. $p$ is the fraction of the surface of the new colloid 
covered by the interaction range of one of its patches. We have chosen $\theta=\pi/6$. The interaction range accounts for both the patch extension and the range of 
the patch-patch interaction. Collisions with the surface always result on the adsorption of the colloids. 

In this model, we consider chemical bonds between patches. This implies that bonds are highly 
directional and therefore promote the alignment of the colloids along their patches. Additionally, chemical bonds are typically characterized by
binding energies larger than the thermal energy. Thus, we assume that binding is irreversible within the timescale of interest.

\subsection{Diffusive Transport}

Diffusion in solution occurs through successive collisions with the solvent. 
We consider particles to diffuse one at a time towards the substrate. The collisions with 
the solvent are described as a Poisson process with a time between collisions, $\Delta t$, exponentially distributed 
  \begin{equation}
   p(\Delta t)=\exp(-R\Delta t),
   \label{eq.time_dist}
  \end{equation}
where $R$ is the collision rate and $p(\Delta t)$ the probability distribution.

At each colloid/solvent collision a new velocity is assigned to the colloid with the direction uniformly distributed and 
the magnitude $v$ following the Maxwell-Boltzmann distribution at the thermostat temperature $T$,
  \begin{equation}
   p(v)=\sqrt{\frac{1}{2\pi T}}\exp\left(-\frac{v^2}{2T}\right),
   \label{eq.maxwell_dist}
  \end{equation}
where $T$ has units of $k_B/m$, $m$ is the mass of the colloid and $k_B$ the Boltzmann constant. 
It is then possible to fine tune the diffusion coefficient $D$ by selecting the adequate thermostat temperature and collision rate.

\subsection{Advective Transport}

To study advective mass transport, we describe the trajectory of the colloids towards the substrate as a straight vertical line (see Fig.~\ref{fig::brownVsball}(b)). 
We randomly generate a position in the horizontal plane and perform a downward vertical movement until the particle either collides with the substrate, sticking to it, 
or to another particle. As in the diffusive case, binding with another particle occurs when their interaction ranges overlap.

\section{Results}\label{sec.results}

First we discuss the case of three-patch colloids and then proceed with the study of mixtures of two- and three-patch colloids.

\subsection{Three-patch colloids}\label{sec.three}

\begin{figure}[t]
\centering
    \includegraphics[width=\columnwidth]{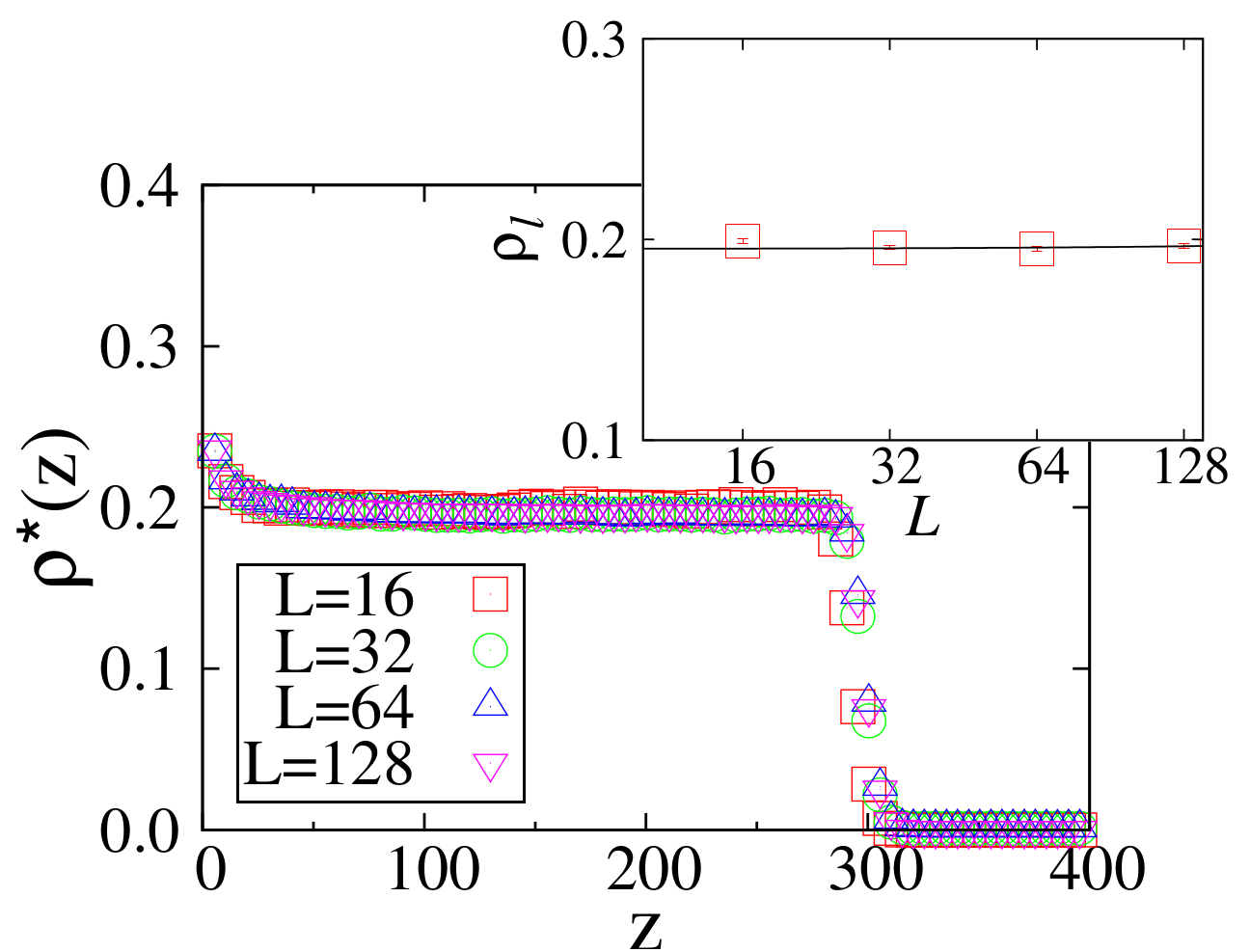} \\
\caption{Density profile of the colloidal network on a substrate, with an advective transport, for values of $L=\{16,32,64,128\}$. 
(inset) Liquid-film density as a function of the substrate lateral size.}
  \label{fig.ballistic_size_nowall}
\end{figure}

\begin{figure}[t]
\centering
    \includegraphics[width=\columnwidth]{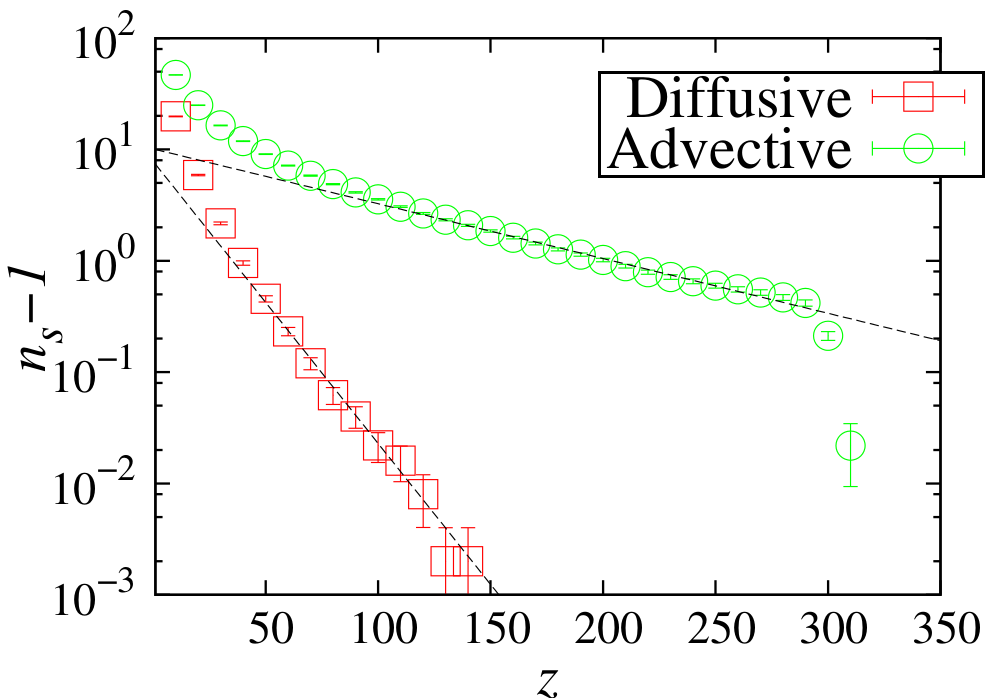} \\
\caption{Number of independent clusters connected to the substrate $n_s$ as a function of the height
 of the film $z$ for diffusive (open squares) and advective (open circles) transports. Measurements were 
 performed for the system of three-patch colloids on a substrate of linear size $L=32$. }
  \label{fig.cluster}
\end{figure}

The aggregation of three-patch colloids on a substrate under diffusive transport has been studied in Ref.\cite{Dias2013}. In that work, 
three regimes in the density profile of the colloidal network were identified: the surface layer, dominated by the influence of the substrate;
the liquid film, where the density of the network of colloids saturates at $\rho=\rho_l$; and the interfacial region, corresponding to the only active region of the structure. 
Figure~\ref{fig.ballistic_size_nowall} shows the density profile for the advective transport of three-patch colloids revealing also three different regimes. 
As we will discuss below, although these regimes are the same as those obtained with diffusion, their structural properties are significantly different. 

In the diffusive case, the structure of the colloidal network is fractal with fractal dimension $d_f=2.58\pm0.04$. This value is consistent with the one 
found for Diffusion Limited Aggregation (DLA) and Diffusion Limited Deposition (DLD)\cite{Meakin1998}. 
The density $\rho_l$ scales as a power law with the lateral size L of the substrate, with the prefactor depending on the diffusion coefficient $D$.
In the advective case, no significant finite-size dependence is observed, as shown in Fig.~\ref{fig.ballistic_size_nowall}.
This result indicates that the colloidal network is not fractal in this case.

\begin{figure}[t]	
  \includegraphics[width=\columnwidth]{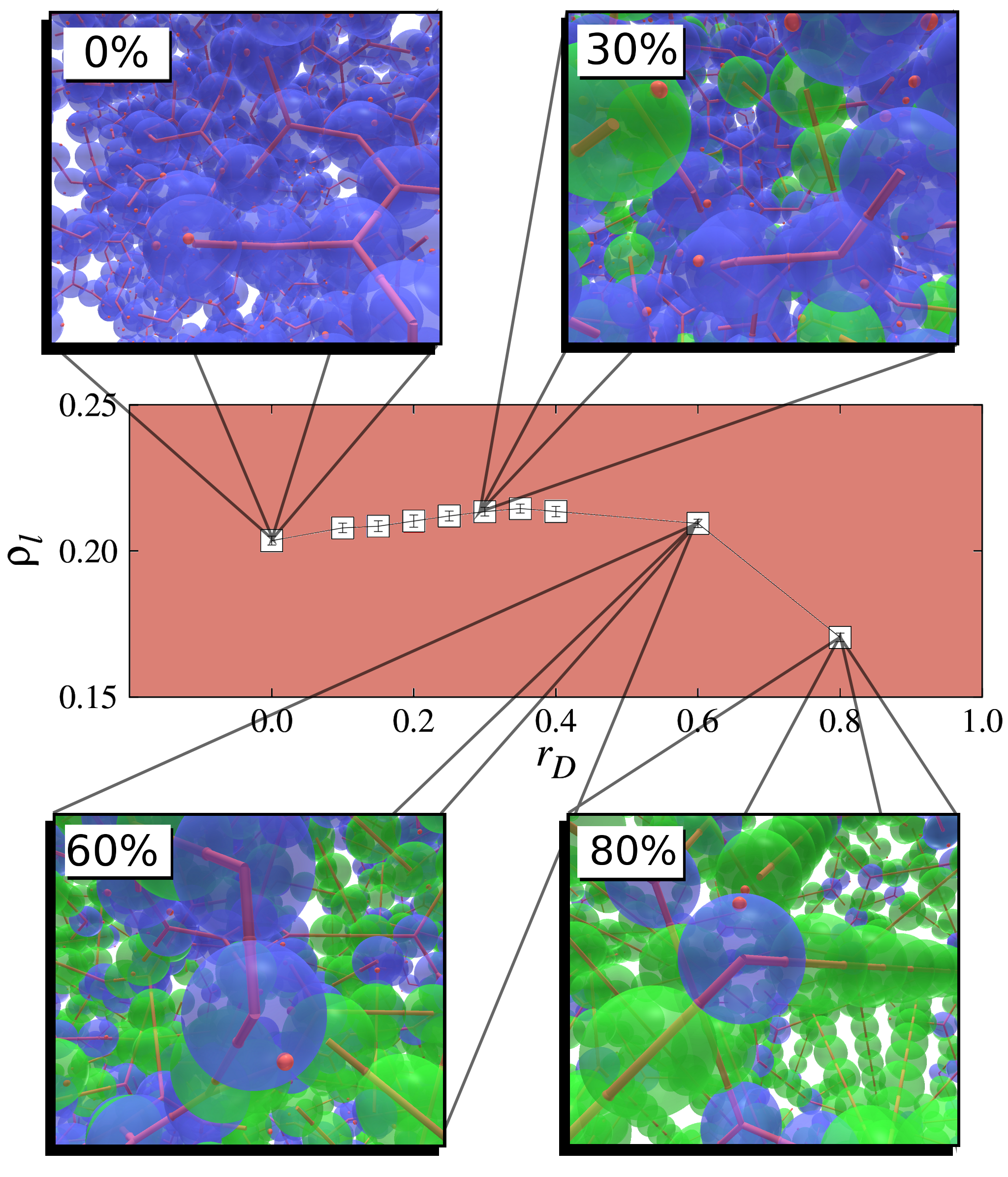}
  \caption{ 
(color online) Density of the liquid regime, for a diffusive transport as a function of the ratio
of two-patch colloids $r_D$ on a substrate of linear size $L=32$. 
Snapshots of a region in the liquid film for different
fractions of two-patch colloids, $r_D$. From left to right, top to
bottom, $r_D$ is $0$, $0.3$, $0.6$, and $0.8$. Three-patch colloids are in blue (dark), two-patch colloids are in
green (light), the (red) spheres on the surface of the colloids represent 
the patches and the (red) sticks the connections between them.
  \label{fig.dimers_snaps_brown}}
\end{figure}

As the colloidal network grows away from the substrate, some branches will grow faster than others and shield the access of new colloids to the inner ones, hindering further binding, 
and the growth of these branches. The competition between branches, however, depends on the mechanism of mass transport. Figure~\ref{fig.cluster} shows the number of 
different branches $n_s$ as a function of the distance $z$ to the substrate. In both cases, the number of branches asymptotically converges towards one and therefore 
we subtract the values in the vertical axis by unity. In both cases, $n_s$ decays exponentially. Nevertheless, the decay is faster under diffusion. The erratic motion 
of diffusing colloids promotes the lateral growth of the network and favors shielding of the inner branches. In the advective case, trajectories are strictly downwards 
and lateral growth is much slower. This is also evident from the largest value of z for advective transport, for the same number of deposited particles.

\begin{figure}[t]	
  \includegraphics[width=\columnwidth]{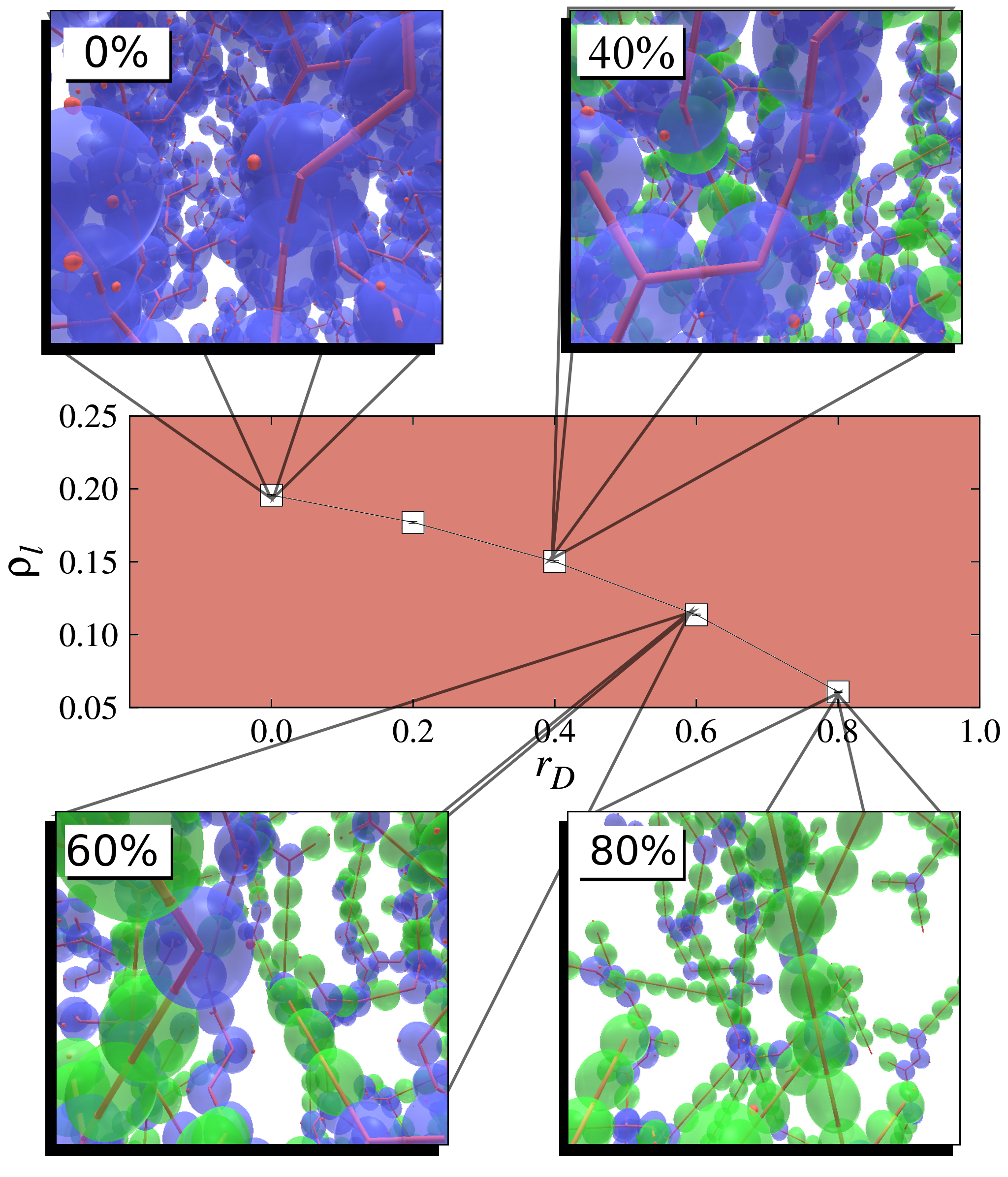}
  \caption{ 
(color online) Density of the liquid regime, for an advective transport as a function of the ratio
of two-patch colloids $r_D$ on a substrate of linear size $L=32$. 
Snapshots of a region in the liquid film for different
fractions of two-patch colloids, $r_D$. From left to right, top to
bottom, $r_D$ is $0$, $0.4$, $0.6$, and $0.8$. Three-patch colloids are in blue (dark), two-patch colloids are in
green (light), the (red) spheres on the surface of the colloids represent 
the patches and the (red) sticks the connections between them.
  \label{fig.dimers_snaps_ball}}
\end{figure}

\begin{figure}[t]	
  \includegraphics[width=\columnwidth]{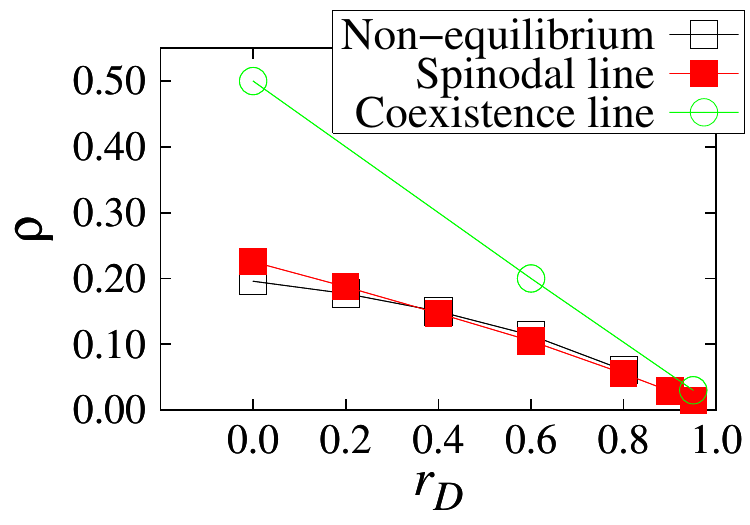}
  \caption{ 
(color online) Density of mixtures of two- and three-patch colloids under equilibrium and nonequilibrium conditions.
Density as a function of the ratio of two-patch colloids for: nonequilibrium liquid film obtained with advective transport (open squares); 
equilibrium spinodal line (solid squares); and equilibrium coexistence curve (open circles). 
Spinodal results from Bianchi et al.\cite{Bianchi2006}. Coexistence results from de las Heras et al.\cite{DelasHeras2011}.
  \label{fig.equilibrium}}
\end{figure}

\subsection{Mixtures of two- and three-patch colloids}\label{sec.two-tree}

When mixtures of two- and three-patch colloids are considered, two mechanisms compete: the formation of long chains favored by two-patch colloids
and branching promoted by three-patch ones. We define the variable $r_D$ as 
the fraction of two-patch colloids. As shown in Fig.~\ref{fig.dimers_snaps_brown}, the larger the $r_D$ the longer the chains 
of two-patch colloids. It was shown in Ref.\cite{Dias2013} that, under diffusive transport, the network is always fractal for any $r_D<1$. 
The measured fractal dimension is consistent with that obtained for three-patch colloids.

In Fig.~\ref{fig.dimers_snaps_brown}, a density maximum is observed at a fraction of two-patch colloids around $0.35$, by contrast to equilibrium 
coexisting gels where a monotonic decrease is observed \cite{Bianchi2006}. This maximum is related to the competition between 
the formation of long chains and the maximization of accessible patches.\cite{Dias2013}

\begin{figure*}[t]	
  \includegraphics[width=0.9\textwidth]{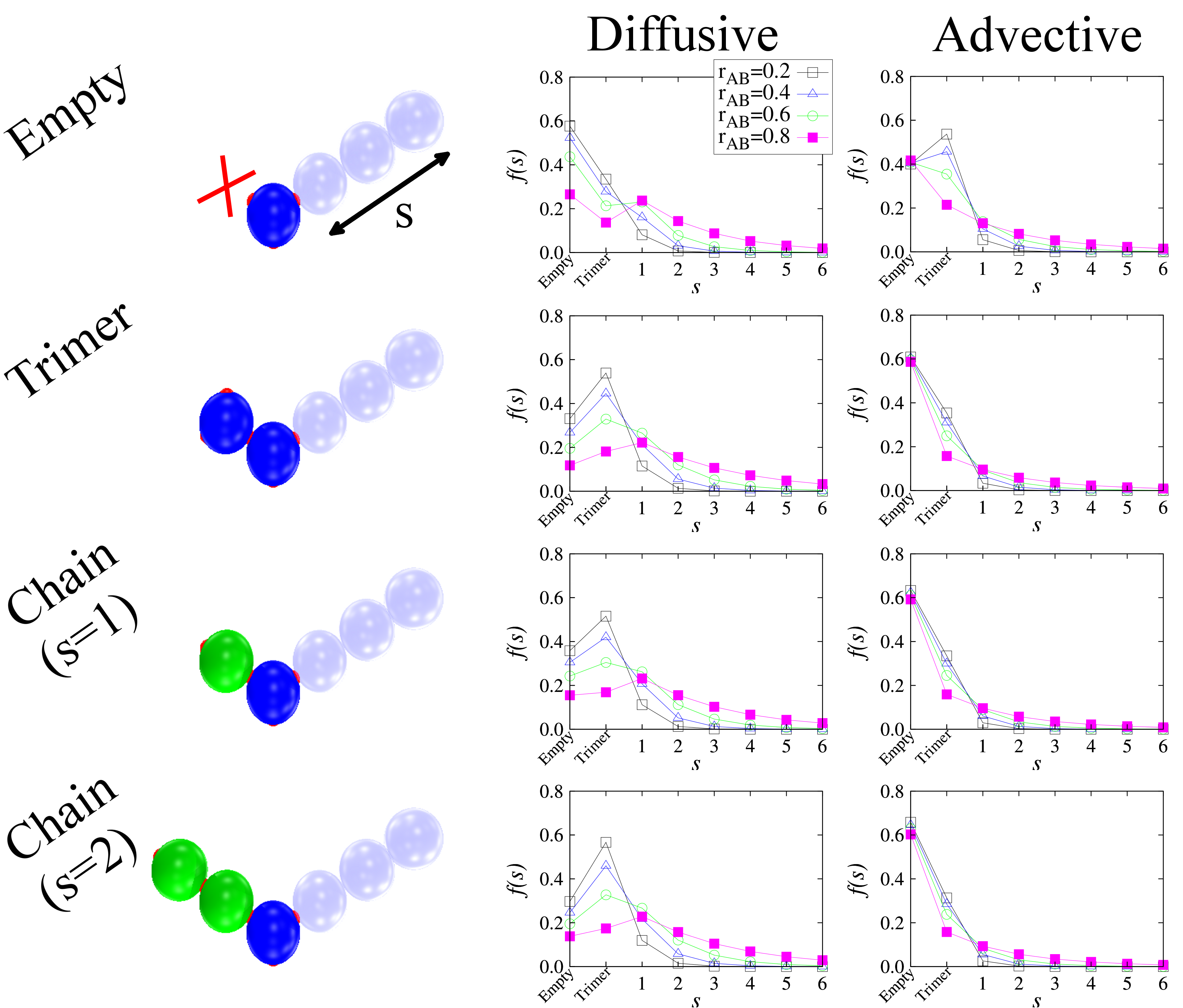}
  \caption{ 
(color online) Chain-growth analysis of three-patch colloids for two mechanisms of transport: diffusive and advective.
(left column) Schematic representation of the configuration of a chain of size $s$ hanging on the neighboring patch of (from top to bottom) an
empty space, a three-patch colloid, a chain of size one, and a chain of size two. (central column) Distribution functions of the chains
for fractions of two-patch colloids of $r_D=\{0.2,0.4,0.6,0.8\}$ for diffusive transport. (right column) Distribution 
functions of the chains for fractions of two-patch colloids $r_D=\{0.2,0.4,0.6,0.8\}$ for advective transport.
  \label{fig.CC_analysis}}
\end{figure*}

The influence of advective transport on the growth of a mixture of two- and three-patch colloids is discussed next. 
The density decreases monotonically with $r_D$, as shown in Fig.~\ref{fig.dimers_snaps_ball}. 
This behavior is similar to that observed under equilibrium.\cite{Bianchi2006} However, in the nonequilibrium case the monotonic decrease is related 
to the mechanism of mass transport. Colloids only follow vertical trajectories towards the substrate and cannot circumvent the branches and squeeze into the fjords 
as in diffusive transport. Therefore, an increase in the fraction of two-patch colloids does not favor the access to available patches in the bulk and only promotes 
the growth of linear chains at the forefront of the network.

In spite of observing similar qualitative behavior for the density in the equilibrium and advective cases, in 
the nonequilibrium case the growth direction (away from the substrate) is imprinted in the branching. 
In Fig.~\ref{fig.equilibrium} we compare the density of the nonequilibrium film with the equilibrium liquid density at coexistence and on the spinodal. 
The coexistence curve corresponds to the region of the phase diagram where liquid and vapor coexist. The spinodal line delimits the region of metastability within the 
coexistence curve. Between the coexistence and the spinodal lines the liquid is mechanically stable.
We note that both equilibrium densities decrease monotonically with $r_D$. Interestingly, the nonequilibrium densities almost overlap the liquid densities along
the spinodal line.
Significant differences are observed only in the limit of small ratio of two-patch colloids, where the density at the spinodal is slightly higher. Note that we 
are comparing the nonequilibrium density on a substrate with equilibrium bulk densities. However, as shown in Ref.\cite{Dias2013}, the properties of the 
network in the liquid film regime are not sensitive to the interaction with the substrate, in line with the behavior of nonequilibrium wetting films.
\cite{DelosSantos2002,DelosSantos2003,DelosSantos2004a}

As described earlier, the maximum in the density observed for diffusive transport is likely driven by the increasing accessibility to empty patches with $r_D$. 
To check this hypothesis we perform a detailed analysis of the film growth mechanisms. The colloidal network grows from several chains 
starting on the substrate which either branch or stop growing at some point. 
Branching is expected to occur at three-patch colloids. 
For each of these particles, in the network, one patch is connected to the chain coming from the substrate. Thus, we measure the correlation 
between the states of the other two patches. We define the state of a patch as: empty, if there is no colloid connected to it; 
trimer, if it is connected to another three-patch colloid; or s-chain, if it is connected to a linear chain of $s$ two-patch colloids.

In the left column of Fig.~\ref{fig.CC_analysis} we plot the distribution of four different configurations. From top to bottom, we plot the size distribution of 
the neighboring chain of an empty patch, a trimer, a 1-chain, and a 2-chain. We found that the distribution of colloids depends on the mechanism of mass transport. 

In diffusive transport, for configurations with an empty patch of reference (first line of Fig.~\ref{fig.CC_analysis}),
there is a higher probability for the neighboring patch to be also empty. This is related to the caging of colloids, as some colloids are trapped 
in cages surrounded by other chains, which hinder the access of new colloids to any of the patches.
Increasing the ratio of two-patch colloids, increases the size of such cages and, thereby, the probability that a diffusing colloid will end up there. For the other 
configurations, the probability of the patch being empty is lower, since, if the colloid is not inside a cage, the diffusive transport allows many 
paths for colloids to bind to both patches.

In advective transport, a qualitatively different picture is observed. If the reference patch is empty, there is a lower probability for the 
neighboring patch to be also empty, when compared with the diffusive case. This probability does not change with the ratio of two-patch colloids since,
regardless of the available volume around the empty patch, a patch that is screened by a chain on top of it will never be reachable during a downward trajectory 
of the particles. For the other configurations, the neighbor patch is mostly empty, as a result of the orientation of the patch and the screening effect. 
This is supported by the distribution of chains for $s\geq1$, where for diffusive transport, 
the larger the chain, the lower the probability of finding an empty patch. By contrast, for advective transport, the probability of finding an
empty patch remains constant for $s=1$ and $s=2$.

\begin{figure}[t]	
  \includegraphics[width=\columnwidth]{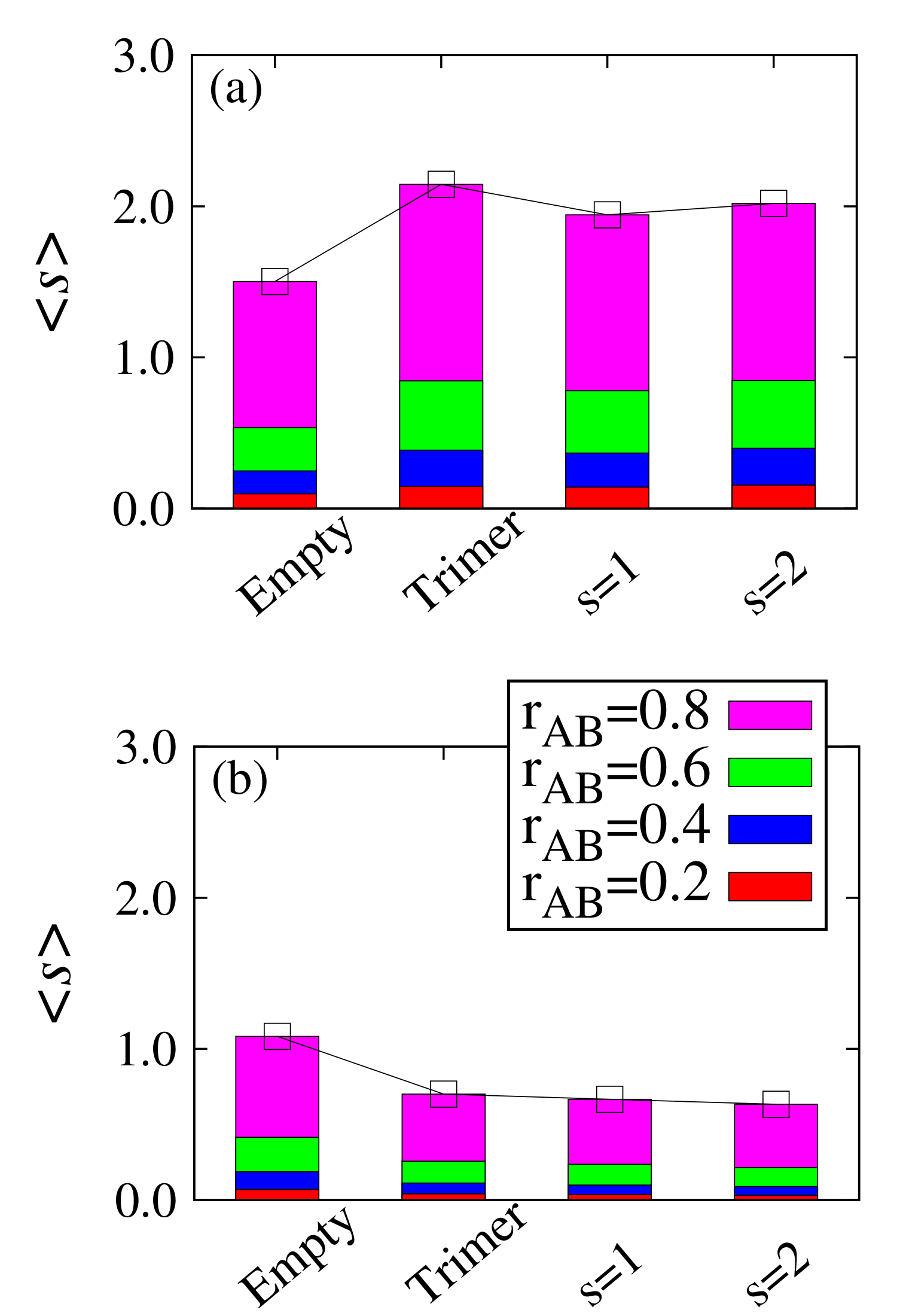}
  \caption{ 
(color online) Average size of two-patch colloid chains near an empty patch, a trimer, or a chain of two-patches colloids of size one and two, 
for two mechanisms of mass transport: (a) diffusion and (b) advection.
  \label{fig.CCA_average}}
\end{figure}

Finally, Fig.~\ref{fig.CCA_average} depicts the average size $s$ of the chains of two-patch colloids. The chains for diffusive transport are typically larger than 
those for the advective case. Note that, for advective transport most of the neighboring patches of an occupied or empty patch are also empty, 
significantly decreasing the average size of the chain (see Fig.~\ref{fig.CC_analysis}).
The reason is that branching implies that both available patches of a three-patch colloid 
are aligned with downward trajectories. This is rarely the case, as the fluctuations in the orientation of the chain and of the 
colloid at the tip will typically shield the patches, hindering branching. The picture is significantly different under diffusion. 
Diffusing particles can circumvent colloids and connect with shielded patches, favoring branching and lateral growth.

\section{Conclusions}\label{sec.conclusions}

We have shown that the kinetics of mass transport towards the substrate significantly affects the structure of the network of patchy particles. 
This is in deep contrast with growth under equilibrium conditions where the history of growth is irrelevant. We observe that, 
diffusive transport favors branching, lateral growth, and the formation of a fractal network of colloids, yielding an optimal 
fraction of two-patch colloids at which the film density is maximized. For advective transport, the network is not fractal and the 
density decreases monotonically with the fraction of two-patch colloids.
Interestingly, we found that the liquid film density, for irreversible adsorption with advective transport, is very 
similar to the one previously observed for the spinodal curve at equilibrium, albeit their structures are quite different. 
The detailed reason for such similarity is an open question, although it may be related to the mechanical instability of the liquid at the spinodal.
Our study highlights the 
dependence of the liquid film structure on the growth conditions. However, we considered the limiting case where binding is irreversible. 
For significantly longer time scales and/or high temperatures, the possibility of bond breaking should 
be considered and relaxation towards equilibrium is to be expected. Future works should consider the dynamics of such relaxation.

\acknowledgments{ We acknowledge financial support from the
Portuguese Foundation for Science and Technology (FCT) under Contracts
nos. EXCL/FIS-NAN/0083/2012, PEst-OE/FIS/UI0618/2011, and
PTDC/FIS/098254/2008. This work was also supported (NA) by grant number FP7-319968 
of the European Research Council.}

\end{document}